\documentstyle[a4,12pt,axodraw]{article}
\newcommand{\ep}{\varepsilon}

\newcommand{\Li}[2]{{\mbox{Li}}_{#1}\left(#2\right)}
\newcommand{\Cl}[2]{{\mbox{Cl}}_{#1}\left(#2\right)}
\newcommand{\Ls}[2]{{\mbox{Ls}}_{#1}\left(#2\right)}
\newcommand{\LS}[3]{{\mbox{Ls}}_{#1}^{(#2)}\left(#3\right)}
\newcommand{\Lsc}[2]{{\mbox{Lsc}}_{#1\!}\left(#2\right)}

\newcommand{\LsLsc}[4]{{\mbox{LsLsc}}_{#1,#2,#3} \left(#4\right)}
\newcommand{\Isc}[2]{{\mbox{Isc}}_{#1}\left(#2 \right)}
\newcommand{\tfrac}[2]{{\textstyle{\frac{#1}{#2}}}}

\newcommand{\Snp}[2]{{\mbox{S}}_{#1\!}\left(#2\right)}

\begin{document}
\begin{flushright}
 hep-ph/0503070 \\[3mm]
 February 2005
\end{flushright}
\begin{flushleft}
{\bf To the memory of  {\it Leonid~Avdeev} }
\end{flushleft}

 \begin{center}
 {\large \bf
About higher order $\ep$-expansion of some massive two- and three-loop 
master-integrals}
 \end{center}
 \vspace*{2.0cm}
\begin{center}
M.~Yu.~Kalmykov
\\
 \vspace{1cm}
Department of Physics, \\
Baylor University, \\
One Bear Place, Box 97316 \\
Waco, TX 76798-7316
\\
\vspace{.3cm}
Bogoliubov Laboratory of Theoretical Physics, \\
Joint Institute for Nuclear Research, \\
$141980$ Dubna (Moscow Region), Russia
\\
\end{center}
 \hspace{3in}
 \begin{abstract}
For certain dimensionally-regulated massive 
two- and three-loop propagator-type diagrams 
the higher order $\ep$-expansion is constructed.
\end{abstract}


\thispagestyle{empty}
\setcounter{page}{0}
\renewcommand{\thefootnote}{\arabic{footnote}}
\setcounter{footnote}{0}

\pagebreak
\section{Introduction}

In the light of the recent progress in the 
four-loop calculations \cite{fourloop,fourloop_2}
one of the topical tasks is the construction 
of $\ep$-expansion of one--, two--,  and three--loop 
master-integrals of the existing packages like 
{\bf ONSHELL2} \cite{onshell2}
or three-loop packages \cite{leo,matad,onshell3,onshell3_2}   
up to level sufficient for getting 
finite four-loop corrections in physical quantities.
The all-order $\ep$-expansion of the one-loop 
propagator diagram with arbitrary masses and 
external momentum, and the two-loop bubble diagram
with arbitrary masses was constructed in 
\cite{D-ep,DK1}. The results are expressible in 
terms of Nielsen polylogarithms \cite{Nielsen} only. 
The construction of the $\ep$-expansion 
for diagrams relating to QED/QCD problems
was investigated in \cite{ON3A,DB96}.
The finite parts of the three-loop bubble 
integrals were presented in \cite{B99,FK99,CS00}.
The higher order $\ep$-expansion for some 
of the master-integrals from these packages 
was calculated  in our previous papers~\footnote{
There is a typo in Eq.~(4.10) of \cite{DK1}: 
the coefficient before $\pi \Ls{4}{\tfrac{\pi}{3}}$ 
should be $\tfrac{161}{54}$ instead of $\tfrac{161}{154}$. 
We are grateful to Y.~Schr\"oder and A.~Vuorinen 
for correspondence.}  \cite{DK1,KV}. 
Independently of analytical calculations, the numerical 
approach to the evaluation of the single scale diagrams 
has been developed in recent years \cite{laporta}.
Based on this technique, the high-precision numerical 
values of higher order coefficients of the $\ep$-expansion 
for four-loop bubble integrals \cite{4loop} and three-loop 
propagator type integrals on mass shell \cite{QED} have 
been calculated recently.

\begin{figure}[th]
\centering
{\vbox{\epsfysize=50mm \epsfbox{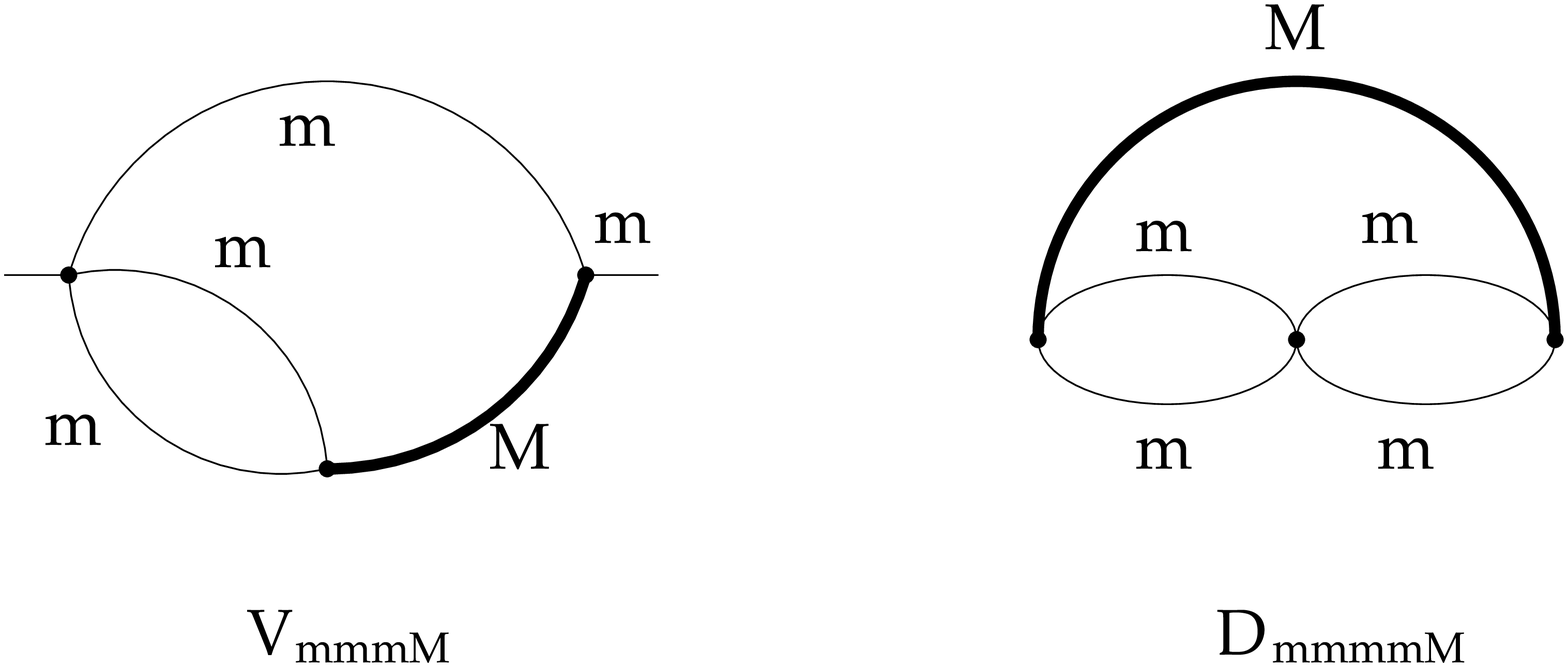}}}
\caption{
}
\label{diagrams}
\end{figure}

The present paper is devoted to the analytical calculation 
of the higher order terms of the $\ep$-expansion  
for the scalar integrals shown in Fig.~\ref{diagrams}.
At $M=m$ these integrals enter in packages 
{\bf ONSHELL2} \cite{onshell2} ({\bf $V_{1111}$})
and three-loop packages \cite{leo,matad} ({\bf $D_5$}).
The numerical values of higher order coefficients 
of the $\ep$-expansion for diagram 
{\bf $V_{1111}$} and {\bf $D_5$}
were calculated by Laporta in \cite{laporta}. 
Another problem under consideration 
is further investigation and development of the ``sixth root of unity'' 
approach proposed by Broadhurst in \cite{B99} and developed in 
\cite{FK99,KV,DK1,DK_tokyo,4loop}.
The main idea is that transcendental numbers 
occurring in the $\ep$-expansion of single scale diagrams (diagrams with only one mass scale)
are defined by the values of massive cuts. For the diagrams with zero--, one--,
two--, and three-- massive cuts, the set of transcendental numbers are related to
generalized log-sine functions \cite{Lewin} or their generalization \cite{DK1,DK_tokyo} 
of special values of arguments 
$\{ \pi, \pi/2, \pi/3, 2\pi/3 \}$. For diagrams with four massive cuts new constants, 
associated with the elliptic function \cite{4loop} appeared. 

We work in the dimensional regularization \cite{dimreg} with space-time dimension $n=4-2\ep$.
In our normalization each loop is
divided by $(4 \pi)^{2-\ep}\Gamma(1+\ep).$ 
We also use the following short notation
for the auxiliary integrals appearing in our calculation:
\begin{eqnarray}
J_{m_1m_2m_3} & = & \left.
\int \frac{d^n k_1 d^n k_2}{[(k_1 \!-\! p)^2 \!+\! m_1^2]
[k_2^2 \!+\! m_2^2][(k_1 \!-\! k_2)^2 \!+\! m_3^3]}
\right|_{p^2=-m^2} \;,
\nonumber \\
V_{m_1m_2m_3} & = & 
\int \frac{d^n k_1 d^n k_2}{[k_1^2 \!+\! m_1^2] [k_2^2 \!+\! m_2^2][(k_1 \!-\! k_2)^2 \!+\! m_3^3]}\;,
\nonumber \\
B_0(m_1,m_2,m_3) & = & \left.
\int \frac{d^n k_1}{[(k_1\!-\!p)^2\!+\!m_1^2] [k_2^2\!+\!m_2^2]} 
\right|_{p^2=-m_3^2} 
\;,
\nonumber \\
A_0(m) & = &   
\int \frac{d^n k_1}{k_1^2 \!+\! m^2}
\equiv \frac{4 m^{n-2}}{(n-2)(n-4)} 
\;.
\label{notations:integrals}
\end{eqnarray}

\section{$V_{mmmM}$}

This integral enters in two-loop relation between pole and $\overline{MS}$ masses 
of heavy particles like t-quark or Higgs boson within SM \cite{top1,top2}.
The integrals with an arbitrary set of indices, 
\begin{equation}
V_{mmmM}(\alpha,\beta,\sigma,\lambda) = 
\int 
\frac{d^nk_1 d^nk_2}{( (p-k_1)^2 + m^2)^{\sigma} ( (k_1-k_2)^2 + m^2)^{\alpha} ( k_2^2 + m^2 )^{\beta}
                    ( k_1^2 + M^2)^{\lambda}}
\end{equation}
where the external moment belongs to the mass shell, $p^2=-m^2$, 
can be reduced to one master-integral with indices (1,1,1,1).
For the calculation of the diagram, we use the differential equation 
technique \cite{diff}
\begin{eqnarray}
&& 
\frac{d}{d M^2} V_{mmmM}(1,1,1,1) = 
V_{mmmM}(1,1,1,1) \Biggl[ 
  \left( \frac{n}{2}-2 \right) \frac{1}{M^2}
+ \frac{(n-3)}{M^2-4m^2}
\Biggr]
\nonumber \\ &&
- \Biggl[ V_{Mmm} + 2 A_0(m) B_0(M,m,m) \Biggr]
\frac{1}{4m^2} \left( \frac{n}{2}-1 \right) \left( \frac{1}{M^2} - \frac{1}{M^2-4m^2} \right)
\nonumber \\ &&
+ J_{mmm}
\frac{1}{m^2} \left(  \frac{3n}{8}-1 \right)  \left( \frac{1}{M^2} - \frac{1}{M^2-4m^2} \right)
\;. 
\label{diff_v1112}
\end{eqnarray}
The analytical solution of this equation up to part linear in $\ep$ was presented 
in \cite{top2}.
Equation (\ref{diff_v1112}) has a very simple form in 
the framework of the geometrical approach \cite{DD}. Let us introduce a new variable
\begin{eqnarray}
\cos\theta = \frac{M}{2m} \; , \quad M \leq 2 m \;. 
\label{angle}
\end{eqnarray}
Taking into account all order $\ep$-expansion of the one-loop propagator 
and two-loop bubble integrals \cite{DT,D-ep}
we rewrite the original differential equation (\ref{diff_v1112}) 
\begin{eqnarray}
&& 
\frac{1}{ \sin \theta \cos \theta} \frac{d}{d \theta} V_{mmmM}
-2 \left( \frac{\ep}{\cos^2 \theta} + \frac{1-2\ep}{\sin^2 \theta} 
   \right) V_{mmmM} 
\nonumber \\ && 
- \frac{1}{2 \cos^2 \theta \sin^2 \theta (1-2\varepsilon)} 
\Biggl\{
- \frac{1}{\ep^2} 
\Biggl[ 
3 
- 6 \cos^2 \theta 
+ 2 \left(4 \cos^2 \theta \right)^{1- \ep}
\Biggr]
\nonumber \\ && 
+ \frac{2 \theta}{\ep} \left(4 \cos \theta \sin \theta \right)^{1-2\ep} 
+  \left(4 \cos \theta \sin \theta \right)^{1-2\ep} 
\sum_{j=0}^{\infty} \frac{(2\varepsilon)^j}{(j+1)!}
\left[ 4 \Ls{j+2}{2 \theta} - 3 \Ls{j+2}{4 \theta} \right]
\Biggr\}
\nonumber \\ && 
+ \frac{J_{mmm}}{m^2} \frac{2-3\ep}{2 \sin^2 \theta \cos^2 \theta }  = 0 \;.
\end{eqnarray}
For a modified function, $\tilde{V}(\theta)$ defined as 
\begin{equation}
V_{mmmM} = 
\left( \sin \theta \right)^{2-4 \ep} \left(\cos \theta \right)^{-2\ep} (m^2)^{-2\ep} \tilde{V}(\theta) \; , 
\end{equation}
the solution has the following form:
\begin{eqnarray}
&& \hspace{-7mm}
2 (1-2\varepsilon)^2 \tilde{V}(\theta) 
=  
- (2 - 3 \ep) (1-2\ep) \frac{J_{mmm}}{(m^2)^{1-2\ep}}
\Biggl[ (1-3\ep) \Isc{4,2}{\theta} 
- \frac{\left( \sin \theta \right)^{4\ep} \left(\cos \theta \right)^{2\ep} } {2 \sin^2 \theta} 
\Biggr] 
\nonumber \\ && \hspace{-5mm}
- \frac{1}{\ep^2} 
\Biggl[  3 (1-\ep) \Isc{4,2}{\theta} 
+ \frac{3}{2} \frac{\left( \sin \theta \right)^{4\ep} \left(\cos \theta \right)^{2\ep} }
               {\sin^2 \theta}
- 4^{1-\ep}  \frac{\left( \sin \theta \right)^{4\ep} }{ \sin^2 \theta} 
\Biggr]
\nonumber \\ && \hspace{-5mm}
+  4^{1-2\ep} \sum_{j=0}^{\infty} \frac{(2\varepsilon)^{j-1}}{j!}
\Biggl\{
- \frac{\cos \theta}{ \left(\sin \theta \right)^{1-2\ep}} 
\left[ 4 \Ls{j+1}{2\theta} - 3 \Ls{j+1}{4 \theta} \right]
\nonumber \\ && \hspace{-5mm}
- 2^{-2 \ep} \sum_{k=0}^\infty  \frac{(2\ep)^k}{k!} 
\left[
8 \frac{\ln^{k+j+1} \left( 2 \sin \theta\right) }{k+j+1}
+ 3 \ep \Ls{k+1}{2 \theta} \Ls{j+1}{4 \theta} 
\right]
\nonumber \\ && \hspace{-5mm}
- 2^{2-2\ep} \ep \sum_{k=0}^\infty  
\frac{(2\ep)^k}{k!} \int^\theta d \phi 
\left[
2 \ln^k \left( 2 \sin \phi \right) \Ls{j+1}{2\phi}
+  3 \Ls{k+1}{2 \phi}
\left[ \ln \left( 2 \sin \phi \right) + \ln \left( 2 \cos \phi \right) \right]^j
\right]
\nonumber \\ && \hspace{-5mm}
+  3 \cdot 4^{1- \ep} \sum_{k=0}^\infty  \frac{(2\ep)^k}{k!} 
\int^\theta d \phi \frac{\cos \phi}{\sin \phi}  \ln^k \left( 2 \sin \phi \right) 
\left[ \ln \left( 2 \sin \phi \right) + \ln \left( 2 \cos \phi \right) \right]^j
\Biggr\} 
\;, 
\label{solution_v1112}
\end{eqnarray}
where $\Ls{j}{\theta}$ is a log-sine function \cite{Lewin} 
and we have introduced a new function $\Isc{a,b}{\theta}$
\begin{equation}
\Isc{a,b}{\theta} = 
\int^\theta
d \phi \frac{ \left(\sin \phi \right)^{a \ep} \left(\cos \phi \right)^{b \ep} }{\sin \phi \cos \phi} \;, 
\label{Isc_ab}
\end{equation}
where $a,b$ are integer numbers. The $\ep$-expansion of this function is 
\begin{eqnarray}
&&  \hspace{-7mm}
\Isc{a,b}{\theta} =  
\ln \left( \tan \theta \right)
+ \frac{1}{2} 
\sum_{i=1}^\infty \frac{ \left( \frac{\ep}{2} \right)^i}{(i+1)!}
\left( a^i \ln^{i+1} \left( \sin^2 \theta \right) 
     - b^i \ln^{i+1} \left( \cos^2 \theta \right) \right) 
\nonumber \\ &&  
- \frac{1}{2} \sum_{i=1}^\infty \left( -\frac{\ep}{2} \right)^i 
\left( a^i \Snp{1,i}{\cos^2 \theta}  - b^i \Snp{1,i}{\sin^2 \theta} \right)
\nonumber \\ &&  
+  \frac{1}{2} \sum_{i,j=1}^\infty a^i b^j \left( \frac{-\ep}{2} \right)^{(i+j)}
\Biggl[
  \sum_{r=0}^{i} \frac{(-1)^r \ln^r \left( \sin^2 \theta \right) }{r!} \Snp{i+1-r,j}{\sin^2 \theta} 
\nonumber \\ &&  \hspace{45mm}
- \sum_{r=0}^{j} \frac{(-1)^r \ln^r \left( \cos^2 \theta \right) }{r!} \Snp{j+1-r,i}{\cos^2 \theta} 
\Biggr] \;,
\label{B_ab}
\end{eqnarray}
where $\Snp{a,b}{\theta}$ is the generalized (Nielsen) polylogarithms \cite{Nielsen}.
The last integral in Eq.~(\ref{solution_v1112}) can be explicitly integrated 
\begin{eqnarray}
&& \hspace{-5mm}
\int^\theta d \theta \frac{\cos \theta}{\sin \theta} \ln^a \left( \sin \theta \right) = 
\frac{1}{1+a} \ln^{a+1} \sin \theta \;, 
\nonumber \\ && \hspace{-5mm}
\int^\theta d \theta \frac{\cos \theta}{\sin \theta} \ln^a  \left( \sin \theta \right) 
                                                    \ln^b  \left( \cos \theta \right)
\!=\! \frac{b! (-1)^b}{2^{1+a+b}} \sum_{i=0}^a \frac{a!(-1)^i}{(a-i)!} 
\left( 2 \ln \left( \sin \theta \right) \right)^{a-i} \Snp{i+1,b}{\sin^2 \theta} \;. 
\label{aux-integral}
\end{eqnarray}
The solution for the original integral can be represented as 
\begin{equation}
V_{mmmM} = 
\left( \sin \theta \right)^{2-4 \ep} \left(\cos \theta \right)^{-2\ep} (m^2)^{-2\ep} 
\Biggl\{
\tilde{V}(\theta)
+ \sum_{j=-1}^\infty \ep^j v_j 
\Biggr\} \;, 
\label{solution_v1112:full}
\end{equation}
where  the coefficients $v_j$ should be defined  from the boundary condition. 
We choose a particular value of the diagram at $M=0$ which corresponds 
to $\theta = \pi/2$.  In this case the diagram reduces to 
$J_{mmm}$ and the product of one-loop vacuum diagrams
$$
\left. V_{mmmM} \right|_{M=0} = 
\frac{3n-8}{4(4-n)} \frac{J_{mmm}}{m^2}
- \frac{3(n-2)^2}{8(n-4)(n-3)} \left[ \frac{A_0(m^2)}{m^2} \right]^2 \; .
$$
Using the results of \cite{ON3A,ON3B} we get the following values 
for the first several coefficients:
\begin{eqnarray}
v_{-1} & = &  12 \ln 2 \;, \quad 
v_0 = 48 \ln 2 + 6 \zeta_2 - 36 \ln^2 2 \;, 
\nonumber \\ 
v_1 & = & 144 \ln 2 + 24 \zeta_2 - 5 \zeta_3 - 12 \zeta_2 \ln 2 - 144 \ln^2 2 + 72 \ln^3 2 \;, 
\nonumber \\ 
v_2 & = &  
384 \ln 2 
\!+\! 72 \zeta_2 
\!-\! 20 \zeta_3 
\!-\! 29 \zeta_4 
\!-\! 48 \zeta_2 \ln 2 
\!+\! 24 \zeta_2 \ln^2 2 
\nonumber \\ && 
- 432 \ln^2 2 
\!+\! 288 \ln^3 2 
\!-\! 106 \ln^4 2 
\!+\! 48 \Li{4}{\tfrac{1}{2}} \;, 
\nonumber \\ 
v_3 & = &  
  960 \ln 2 
\!+\! 192 \zeta_2 
\!-\! 60 \zeta_3 
\!-\! 116 \zeta_4 
\!-\! \tfrac{633}{2} \zeta_5
\!-\! 144 \zeta_2 \ln 2
\!+\! 96 \zeta_2 \ln^2 2
\nonumber \\ && 
\!-\! 48 \zeta_2 \ln^3 2
\!+\! 306 \zeta_4 \ln 2
\!-\! 1152 \ln^2 2
\!+\! 864 \ln^3 2
\!-\! 424 \ln^4 2
\!+\! \tfrac{636}{5} \ln^5 2
\nonumber \\ && 
\!+\! 192 \Li{4}{\tfrac{1}{2}} 
\!+\! 288 \Li{5}{\tfrac{1}{2}} \; .
\end{eqnarray}
$v_4$ can also be calculated by using the results of \cite{ON3B}.

From Eq.~(\ref{solution_v1112})
it is evident, that the coefficients of the $\ep$-expansion of the
diagram $V_{1112}$ can be parametrized by introducing
a new class of functions $\LsLsc{k}{i}{j}{\theta}$
defined as 
\begin{eqnarray}
\LsLsc{k}{i}{j}{\theta} = \int_0^\theta d \phi \Ls{k+1}{\phi}
\ln^{i-1} \left| 2 \sin  \frac{\phi}{2}  \right| 
\ln^{j-1} \left| 2 \cos  \frac{\phi}{2}  \right| \; , 
\end{eqnarray}
where $k,i,j$ are integer numbers, $k \geq 0$ and $i,j \geq 1$. 
Some properties of these functions are collected 
in Appendix~\ref{LsLogsincos}.

Let us write the $\ep$-expansion of the diagram in following form:
\begin{eqnarray}
V_{mmmM}(\theta) = 
(m^2)^{-2 \ep} \Biggl[
\frac{1}{2 \ep^2}
\!+\! \frac{1}{\ep } S(\theta)
\!+\! F(\theta)
\!+\! \ep E(\theta)
\!+\! \ep^2 N(\theta)
\!+\! \ep^3 N_n(\theta)
\!+\! \ep^4 N_{nn}(\theta)
\!+\! {\cal O}(\ep^5)
\Biggr] 
\;.
\nonumber \\ 
\end{eqnarray}
The functions $S(\theta), F(\theta)$ and $E(\theta)$ were calculated in 
\cite{top2} (see Eq.(3.27)). 
The analytical results for $N(\theta)$ for arbitrary values of the angle $\theta$ 
read
\begin{eqnarray}
&& 
\frac{N(\theta)}{\sin^2 \theta} = 
4 \frac{F(\theta) - E(\theta)}{\sin^2 \theta}
- 4 \theta  \biggl[ 4 \Ls{3}{2\theta} - 3 \Ls{3}{4\theta} \biggr]
\nonumber \\ && 
- \frac{2}{3} \frac{\sin (2 \theta) }{\sin^2 \theta}
\sum_{j=0}^3  (-1)^j \biggl[4 \Ls{4-j}{2\theta} - 3 \Ls{4-j}{4\theta} \biggr] l^j_{2 \theta} 
\left( 3 \atop j \right)
\nonumber \\ && 
+ \frac{1}{\sin^2 \theta}
\Biggl\{
  \tfrac{27}{2}
+ \tfrac{4}{3} \cos^2 \theta L_\theta^4
+ 24 \zeta_2 \ln 2 
- 14 \zeta_3 
- 12 \zeta_2 
\Biggr \}
\nonumber \\ && 
- 24 \Phi( 2 \theta) 
+ 12 \LS{4}{1}{2 \theta}
- 3  \LS{4}{1}{4 \theta}
+ 20 \Biggl[ \Ls{2}{2 \theta} \Biggr]^2
- 12 \Ls{2}{2 \theta} \Ls{2}{4 \theta}
\nonumber \\ && 
- 16 \theta L_\theta \Ls{2}{2 \theta}
- 32 \theta l_\theta \Ls{2}{2 \theta}
+ 24 \Cl{3}{\pi - 2 \theta} \left[ L_\theta + 2 l_\theta \right]
- 12 \zeta_2 \Li{2}{\sin^2 \theta}
\nonumber \\ && 
- 8 \theta^2 L^2_\theta
- 32 \theta^2 l_\theta^2 
- 32 \theta^2 L_\theta l_\theta 
+ 36 \zeta_3 l_\theta
- 24 \zeta_3 L_\theta
+ 12 \zeta_2 L_\theta^2
+ 48 \ln 2 \zeta_2 L_\theta
\nonumber \\ && 
+ 42 \zeta_3 \ln 2 
- 31 \zeta_4 
+ 2 \ln^4 2 
- 36 \zeta_2 \ln^2 2 
+ 48 \Li{4}{\tfrac{1}{2}} \; , 
\end{eqnarray}
where 
\begin{equation}
L_\theta = \ln \left(2 \cos \theta \right) \; , \quad 
l_\theta = \ln \left(2 \sin \theta \right) \; , \quad 
l_{m \theta} = \ln \left(2 \sin m \theta \right) \; ,
\label{Ll}
\end{equation}
and 
$$
\Phi(\theta) \equiv \int_0^\theta d \phi \Ls{2}{\phi} \ln \left(2 \cos \frac{\phi}{2} \right) 
= \LsLsc{1}{1}{2}{\theta} \;,
$$
is the function defined in Eq.~(2.41) of \cite{DK2}.
The results for $N_n(\theta)$ and $N_{nn}(\theta)$ are
sufficiently lengthy to be published here.
It should be mentioned that in the order of $\ep^3$ 
the following combination appears:
\begin{eqnarray}
\hspace{-3mm}
2 \LsLsc{1}{2}{2}{2 \theta}
\!+\!
\LsLsc{0}{3}{2}{2 \theta} \!+\! \LsLsc{0}{3}{2}{\pi \!-\! 2 \theta} 
\!+\!
\LsLsc{1}{1}{3}{2 \theta} \!-\! \LsLsc{1}{1}{3}{\pi \!-\! 2 \theta} \; . 
\label{combination}
\end{eqnarray}
To write results valid in other regions of the variable $(M > 2m)$, 
a proper analytical continuation of all expressions 
should be constructed. For generalized log-sine functions it is 
described in detail in \cite{DK1,DK_bastei}. 
In terms of the variable 
\begin{eqnarray}
y \equiv e^{ {\rm i} \sigma 2 \theta}, \hspace{5mm}
\ln(-y-{\rm i}\sigma 0) = \ln{y} - {\rm i} \sigma \pi,
\label{y}
\end{eqnarray}
the analytical continuation of all generalized log-sine can be expressed 
in terms of Nielsen polylogarithms, whereas for the function $\Phi(\theta)$
the result is expressible in terms of harmonic polylogarithms 
(see Eq.(3.4) in \cite{DK2}) introduced by Remiddi and Vermaseren \cite{RV00}.
In terms of conformal variable (\ref{y}), the analytical continuation of the 
${\rm LsLsc}$ functions can be written in terms of harmonic polylogarithms 
(see the discussion in \cite{DK1,DK2,MK2004}). In particular, an analytical 
continuation of the $\LsLsc{1}{1}{3}{\theta}$-function produces integral 
$\int H_{-1,0,0,1}(z) \tfrac{dz}{1+z}$.
The result is relatively lengthy to be published here.
The analytically continued results of the order $\ep^2$
checked by heavy mass expansion~\cite{asymptotic} 
with the help of the packages described in~\cite{tlamm}.

For a particular case $M=m$ the integral $V_{mmmM}$ converts into the master-integral $V_{1111}$
from the package {\bf ONSHELL2} \cite{onshell2,single}. Its $\ep$-expansion is (we present only new 
coefficients of the expansion)
\begin{eqnarray}
&& 
N \left( \tfrac{\pi}{3} \right) \equiv 
V_{1111}[\ep^2]  = 
\tfrac{211}{2}
- 32 \tfrac{\pi}{\sqrt{3}}
+ 12 \tfrac{\pi}{\sqrt{3}} \ln 3 
- 2  \tfrac{\pi}{\sqrt{3}} \ln^2 3 
+ \tfrac{1}{6} \tfrac{\pi}{\sqrt{3}} \ln^3 3 
\nonumber \\ && 
- 34 \zeta_2 
+ 16 \zeta_2 \ln 3 
- 4 \zeta_2 \ln^2 3 
- 32 \zeta_3 
+ \tfrac{39}{2} \zeta_3 \ln 3 
- 18 \zeta_2  \tfrac{\pi}{\sqrt{3}} 
+ \tfrac{9}{2} \zeta_2 \tfrac{\pi}{\sqrt{3}} \ln 3  
+ 9 \zeta_3 \tfrac{\pi}{\sqrt{3}} 
\nonumber \\ && 
- 84 \tfrac{\Ls{2}{\tfrac{\pi}{3}}}{\sqrt{3}}
+ 28 \tfrac{\Ls{2}{\tfrac{\pi}{3}}}{\sqrt{3}} \ln 3 
+ \tfrac{16}{3} \pi \Ls{2}{\tfrac{\pi}{3}}
- 42 \tfrac{\Ls{3}{\tfrac{2 \pi}{3}}}{\sqrt{3}}
\nonumber \\ && 
- \tfrac{7}{2} \tfrac{\Ls{2}{\tfrac{\pi}{3}}}{\sqrt{3}} \ln^2 3 
- \tfrac{8}{3} \pi \Ls{2}{\tfrac{\pi}{3}} \ln 3 
+ \tfrac{21}{2} \tfrac{\Ls{3}{\tfrac{2 \pi}{3}}}{\sqrt{3}} \ln 3 
- 7 \pi \Ls{3}{\tfrac{2 \pi}{3}}
\nonumber \\ && 
- \tfrac{219}{4} \zeta_4 
- 7 \tfrac{\Ls{4}{\tfrac{2 \pi}{3}}}{\sqrt{3}}
+ \tfrac{27}{2} \LS{4}{1}{\tfrac{2 \pi}{3}}
+ \tfrac{14}{3} \Biggl[ \Ls{2}{\tfrac{\pi}{3}} \Biggr]^2
\nonumber \\ && 
+ 24 \zeta_2 \ln 2 
+ 18 \zeta_2 \ln^2 2 
- 18 \zeta_2 \ln 2 \ln 3 
+ 9 \zeta_2 \Li{2}{\tfrac{1}{4}} \;,
\label{v1111_ep^2}
\end{eqnarray}
and
\begin{eqnarray}
&& \hspace{-5mm}
N_n \left( \tfrac{\pi}{3} \right) \equiv 
V_{1111}[\ep^3]  = 
\tfrac{665}{2}
- 80 \tfrac{\pi}{\sqrt{3}}
+ 32 \tfrac{\pi}{\sqrt{3}} \ln 3 
- 6 \tfrac{\pi}{\sqrt{3}} \ln^2 3 
+ \tfrac{2}{3}  \tfrac{\pi}{\sqrt{3}} \ln^3 3 
- \tfrac{1}{24} \tfrac{\pi}{\sqrt{3}}\ln^4 3 
\nonumber \\ && \hspace{-5mm}
- 54 \zeta_2 \tfrac{\pi}{\sqrt3} 
+ 18 \zeta_2 \tfrac{\pi}{\sqrt3} \ln 3 
- \tfrac{9}{4} \zeta_2 \tfrac{\pi}{\sqrt3} \ln^2 3 
- 148 \zeta_2 
- 152 \zeta_3 
- 157 \zeta_4 
+ \tfrac{1081}{24} \zeta_2 \zeta_3 
+ \tfrac{3593}{12} \zeta_5
\nonumber \\ && \hspace{-5mm}
+ 48 \zeta_2 \ln 3 
- 16 \zeta_2 \ln^2 3
+ \tfrac{8}{3} \zeta_2 \ln^3 3 
+ 36 \zeta_3 \tfrac{\pi}{\sqrt3} 
- 9 \zeta_3  \tfrac{\pi}{\sqrt3} \ln 3 
+ 78 \zeta_3 \ln 3 
- \tfrac{39}{2} \zeta_3 \ln^2 3 
\nonumber \\ && \hspace{-5mm}
+ 63 \zeta_4 \ln 3 
- \tfrac{171}{8} \zeta_4 \tfrac{\pi}{\sqrt3} 
- 224 \tfrac{\Ls{2}{\tfrac{\pi}{3}}}{\sqrt{3}}
+ 84 \tfrac{\Ls{2}{\tfrac{\pi}{3}}}{\sqrt{3}} \ln 3
- 14 \tfrac{\Ls{2}{\tfrac{\pi}{3}}}{\sqrt{3}} \ln^2 3
+ \tfrac{7}{6} \tfrac{\Ls{2}{\tfrac{\pi}{3}}}{\sqrt{3}} \ln^3 3
\nonumber \\ && \hspace{-5mm}
+ 16 \pi \Ls{2}{\tfrac{\pi}{3}} 
- \tfrac{32}{3} \pi \Ls{2}{\tfrac{\pi}{3}} \ln 3 
+ \tfrac{8}{3} \pi \Ls{2}{\tfrac{\pi}{3}} \ln^2 3 
+ 6 \pi \zeta_2 \Ls{2}{\tfrac{\pi}{3}}
\nonumber \\ && \hspace{-5mm}
- 126 \tfrac{\Ls{3}{\tfrac{2 \pi}{3}}}{\sqrt{3}} 
+ 42 \tfrac{\Ls{3}{\tfrac{2 \pi}{3}}}{\sqrt{3}} \ln 3 
- \tfrac{21}{4} \tfrac{\Ls{3}{\tfrac{2 \pi}{3}}}{\sqrt{3}} \ln^2 3 
- 14 \pi \Ls{3}{\tfrac{2 \pi}{3}} \left[ 2 - \ln 3 \right]
\nonumber \\ && \hspace{-5mm}
+ \tfrac{23}{18} \pi \Ls{4}{\tfrac{\pi}{3}}
- \tfrac{23}{3} \pi \Ls{4}{\tfrac{2\pi}{3}}
- 7 \tfrac{\Ls{4}{\tfrac{2 \pi}{3}}}{\sqrt{3}} \left[ 4 - \ln 3 \right]
+ 27 \LS{4}{1}{\tfrac{2\pi}{3}} \left[ 2 - \ln 3 \right]
\nonumber \\ && \hspace{-5mm}
- \tfrac{7}{2} \tfrac{\Ls{5}{\tfrac{2 \pi}{3}}}{\sqrt{3}}
+ \tfrac{28}{3} \left[\Ls{2}{\tfrac{\pi}{3}} \right]^2 \left[ 2 - \ln 3 \right]
- 4 \Ls{2}{\tfrac{\pi}{3}} \Ls{3}{\tfrac{2\pi}{3}}
- \tfrac{55}{6} \pi \zeta_2 \Ls{2}{\tfrac{\pi}{3}}
\nonumber \\ && \hspace{-5mm}
+ 9 \LS{5}{1}{\tfrac{2\pi}{3}}
- \tfrac{9}{8} \chi_5
+ 36 \LsLsc{1}{1}{3}{\tfrac{\pi}{3}}
+ 168 \zeta_2 \ln 2 
\nonumber \\ && \hspace{-5mm}
- 4 \ln^4 2 
+ 24 \zeta_2 \ln^2 2 
- 96 \Li{4}{\tfrac{1}{2}}
- \tfrac{48}{5} \ln^5 2 
- 63 \zeta_3 \ln^2 2 
- 288 \left[ \Li{4}{\tfrac{1}{2}} \ln 2  + \Li{5}{\tfrac{1}{2}} \right]
\nonumber \\ && \hspace{-5mm}
+ \Li{2}{\tfrac{1}{4}} 
\Biggl[
  36 \zeta_2 
+ \tfrac{63}{2} \zeta_3 
- 18 \zeta_2 \ln 2 
- 18 \zeta_2 \ln 3 
\Biggr]
- 9 \zeta_2 \Li{3}{\tfrac{1}{4}}
- 18 \zeta_2 \Snp{1,2}{\tfrac{1}{4}}
\nonumber \\ && \hspace{-5mm}
+ 72  \Li{4}{\tfrac{1}{2}} \ln 3 
+ 3 \ln 3 \ln^4 2 
- 72 \zeta_2 \ln 2 \ln 3 
+ 18 \zeta_2 \ln 2 \ln^2 3 
\; .
\label{v1111_ep^3}
\end{eqnarray}
Numerical values of the coefficients, $V_{1111}[\ep^2]$
and $V_{1111}[\ep^3]$ are in full agreement with the proper 
results of \cite{laporta}.
\section{$D_{mmmmM}$}

The differential equation for this integral is 
\begin{eqnarray}
&& 
- M^2 \frac{d}{d M^2} D_{mmmmM}
= \left( 5 +  4 (3-n) \frac{m^2}{M^2-4m^2} - \frac{3}{2} n \right) D_{mmmmM}
\nonumber \\ && 
+ B_N \left(\frac{3}{2} n - 4 \right) \frac{1}{M^2-4m^2}
+ V_{Mmm} A_0(m) \frac{4-2n}{M^2-4m^2} \; , 
\label{diff_d}
\end{eqnarray}
where $B_N \equiv B_N(0,0,1,1,1,1)$ was defined by Broadhurst in \cite{ON3A,DB96}.
Using again the angle variable defined in (\ref{angle}), we rewrite equation (\ref{diff_d})
in the following form:
\begin{eqnarray}
&& 
\frac{\cos \theta}{ 2 \sin \theta} \frac{d }{d \theta} D_{mmmmM}
- \left( -1 + 3 \ep  + \frac{1-2\ep}{\sin^2 \theta} \right) D_{mmmmM}
\nonumber \\ && 
= 
- \frac{B_N}{m^2} \frac{(2-3\ep)}{4} \frac{1}{\sin^2 \theta} 
- \frac{(m^2)^{1-3\ep}}{\sin^2 \theta } \frac{1}{\ep(1-\ep)(1-2\ep)}
\Biggl\{
- \frac{1}{\ep^2} 
\Biggl[ 
1
- 2 \cos^2 \theta 
+ \left(4 \cos^2 \theta \right)^{1- \ep}
\Biggr]
\nonumber \\ && 
+  \left(4 \cos \theta \sin \theta \right)^{1-2\ep} 
\sum_{j=0}^{\infty} \frac{(2\varepsilon)^j}{(j+1)!}
\left[ 2 \Ls{j+2}{2 \theta} - \Ls{j+2}{4 \theta} \right]
\Biggr\} \; , 
\end{eqnarray}

For auxiliary function $\tilde{D}(\theta)$ defined as 
$$
D_{mmmmM} = 
\left( \sin \theta \right)^{2-4 \ep} \left(\cos \theta \right)^{-2\ep} (m^2)^{1-3\ep} \tilde{D}(\theta)
$$
the solution is 
\begin{eqnarray}
&& \hspace{-7mm}
(1-\ep) (1-2\ep)^2 \tilde{D}(\theta) = 
\nonumber \\ && \hspace{-5mm}
- \frac{(1\!-\!\ep) (2\!-\!3\ep) (1\!-\!2\ep)}{2}  \frac{B_N}{(m^2)^{2-3\ep}}
\left[  (1-3\ep) \Isc{4,2}{\theta} 
-  \frac{ \left( \sin \theta \right)^{4\ep} \left(\cos \theta \right)^{2\ep} } {2 \sin^2 \theta }
\right] 
\nonumber \\ && \hspace{-5mm}
+ \frac{2}{\ep^3} 
\Biggl[
  (1-\ep) \Isc{4,2}{\theta} 
+ \frac{\left( \sin \theta \right)^{4\ep} \left(\cos \theta \right)^{2\ep} } {2 \sin^2 \theta}
- 2^{1-2\ep}  \frac{\left( \sin \theta \right)^{4\ep} }{ \sin^2 \theta} 
\Biggr]
\nonumber \\ && \hspace{-5mm} 
- 2^{3-4\ep} \frac{1}{\ep} 
\sum_{j=0}^{\infty} \frac{(2\varepsilon)^{j}}{(j+1)!}
\Biggl\{
- \frac{\cos \theta}{ \left(\sin \theta \right)^{1-2\ep}}
\left[ 2 \Ls{j+2}{2\theta} - \Ls{j+2}{4 \theta} \right]
\nonumber \\ && \hspace{-5mm}
- 2^{-2 \ep} \sum_{k=0}^\infty \frac{(2\ep)^k}{k!} 
\left[ 
4 \frac{\ln^{k+j+2} \left( 2 \sin \theta\right) }{k+j+2}
+ \ep  \Ls{k+1}{2 \theta} \Ls{j+2}{4 \theta} 
\right]
\nonumber \\ && \hspace{-5mm}
- 2^{2-2\ep} \ep \sum_{k=0}^\infty 
\frac{(2\ep)^k}{k!} \int^\theta d \phi 
\left[ 
\Ls{k+1}{2 \phi}
\left[ \ln \left( 2 \sin \phi \right) + \ln \left( 2 \cos \phi \right) \right]^{j+1}
+ \ln^k \left( 2 \sin \phi \right) \Ls{j+2}{2\phi}
\right]
\nonumber \\ && \hspace{-5mm}
+  2^{2- 2\ep} \sum_{k=0}^\infty \frac{(2\ep)^k}{k!} 
\int^\theta d \phi \frac{\cos \phi}{\sin \phi}  \ln^k \left( 2 \sin \phi \right) 
\left[ \ln \left( 2 \sin \phi \right) + \ln \left( 2 \cos \phi \right) \right]^{j+1}
\Biggr\}
\;, 
\label{solution_d11112}
\end{eqnarray}
where the last integral can be explicitly integrated with the  help of Eq.~(\ref{aux-integral}) and 
$\Isc{a,b}{\theta}$ is defined in (\ref{B_ab}).
The result for the diagram $D_{mmmmM}$ is 
\begin{equation}
D_{mmmmM} = 
\left( \sin \theta \right)^{2-4 \ep} \left(\cos \theta \right)^{-2\ep} (m^2)^{1-3\ep} 
\Biggl\{ \tilde{D}(\theta) + \sum_{j=-3}^\infty \ep^j d_j \Biggr\} \;, 
\end{equation}
where  the coefficients $d_j$ are defined  from the value of the original 
diagrams at $M=0$, which corresponds to $\theta = \pi/2$:
\begin{eqnarray}
\left. m^2 D_{mmmmM} \right|_{M=0} = V_{0mm} A_0(m) \frac{n-2}{n-4} - B_N \frac{3n-8}{4(n-4)} \; . 
\end{eqnarray}
The explicit values of the first several coefficients are 
\begin{eqnarray}
d_{-3} & = &  \tfrac{4}{3} \;, \quad 
d_{-2}  =  \tfrac{20}{3} - 8 \ln 2  \;, \quad 
d_{-1}  =   \tfrac{68}{3} - \tfrac{16}{3} \zeta_2  - 40 \ln 2  - 8 \ln^2 2 \;, 
\nonumber \\ 
d_0 & = &  \tfrac{196}{3} + \tfrac{20}{3} \zeta_3 - \tfrac{80}{3} \zeta_2 
           - 136 \ln2  - 40 \ln^2 2  + \tfrac{176}{3} \ln^3 2 \;, 
\nonumber \\ 
d_1 & = &   - \tfrac{400}{3} \ln^4 2  + \tfrac{880}{3} \ln^3 2 + 16 \zeta_2 \ln^2 2 -  136 \ln^2 2
            - 392 \ln 2  
\nonumber \\ && 
- \tfrac{272}{3} \zeta_2  + \tfrac{100}{3} \zeta_3  + \tfrac{116}{3} \zeta_4  
            - 64 \Li{4}{\tfrac{1}{2}} + 172 \;, 
\nonumber \\ 
d_2 & = &  
  428 
+ \tfrac{2912}{15} \ln^5 2 
- \tfrac{2000}{3} \ln^4 2 
+ \tfrac{2992}{3} \ln^3 2 
- 392 \ln^2 2
- 1032 \ln 2 
\nonumber \\ && 
- 32 \zeta_2 \ln^3 2  
+ 80 \zeta_2 \ln^2 2 
- 408 \zeta_4 \ln 2 
- \tfrac{784}{3} \zeta_2 
+ \tfrac{340}{3} \zeta_3 
+ \tfrac{580}{3} \zeta_4 
\nonumber \\ && 
+ \tfrac{16}{3} \zeta_2 \zeta_3 
+ 422 \zeta_5 
- 320 \Li{4}{\tfrac{1}{2}} 
- 384 \Li{5}{\tfrac{1}{2}} 
\;, 
\end{eqnarray}
The next coefficient $d_3$ contains the $U_{5,1}$ constant \cite{DB96} which could be rewritten 
in terms of the generalized log-sine function of the argument $\pi/2$ \cite{DK1}. 
Collecting all the results we get the first several coefficients of 
the $\ep$-expansion of the diagram
\begin{eqnarray}
&& \hspace{-7mm}
(m^2)^{-1+3\ep}
(1-\ep)(1-2\ep)^2 D_{mmmmM}(\theta) = 
- \frac{2}{3\ep^3}
\left\{ 1 + 2 \cos^2 \theta \right\}
- \frac{2}{3 \ep^2}
\left\{ 1 - 12 \cos^2 \theta L_{\theta} \right\}
\nonumber \\ &&  \hspace{-5mm}
- \frac{2}{\ep}
\left\{ 1 + 4 \cos^2 \theta L^2_{\theta}
- 2 \sin( 2 \theta) \left[ 2 \Ls{2}{2 \theta} - \Ls{2}{4 \theta} \right]
\right\}
\nonumber \\ && \hspace{0mm}
- 6
+ \frac{16}{3} \cos^2 \theta L^3_{\theta}
- 4 \sin^2 \theta 
\left[ 
4 \Cl{3}{2 \theta}
- \Cl{3}{4 \theta}
- \frac{2}{3} \zeta_3 
\right]
\nonumber \\ && \hspace{0mm}
+ 4 \sin (2 \theta)
\left[ 2 \Ls{3}{2 \theta} - \Ls{3}{4 \theta} \right]
- 8 \sin (2 \theta) l_{2 \theta}
\left[ 2 \Ls{2}{2 \theta} - \Ls{2}{4 \theta} \right]
\nonumber \\ &&  \hspace{-5mm}
+ \ep \Biggl\{ 
- 18 
+ \frac{56}{3} \zeta_3 
+ 8 \sin^2 \theta 
\left[2 l_\theta  + L_\theta \right]
\left[ 4 \Cl{3}{2 \theta} - \Cl{3}{4 \theta} \right]
\nonumber \\ && \hspace{2mm}
+ 4 \sin^2 \theta 
\left[
  8 F(2 \theta)
- 4 \LS{4}{1}{2 \theta}
+   \LS{4}{1}{4 \theta}
\right]
\nonumber \\ && \hspace{2mm}
+ 16 \sin^2 \theta 
\Biggl[
\theta \left[ 2 \Ls{3}{2 \theta} \!-\! \Ls{3}{4 \theta} \right]
\!-\! \Ls{2}{2\theta} \left[ 2 \Ls{2}{2 \theta} \!-\! \Ls{2}{4 \theta} \right]
\!+\! \zeta_3 \left( 2 L_{\theta} \!-\! 3 l_{\theta} \right)
\Biggr]
\nonumber \\ &&  \hspace{2mm}
+ 4 \sin^2 \theta 
\left[
17 \zeta_4 
- 16 \Li{4}{\tfrac{1}{2}}
+ 4 \zeta_2 \ln^2 2 
- 14 \zeta_3 \ln 2 
- \frac{2}{3} \ln^4 2 
\right]
- \frac{8}{3} \cos^2 \theta L^4_{\theta}
\nonumber \\ && \hspace{2mm}
+ 8 \sin (2 \theta) l^2_{2 \theta}
\left[ 2 \Ls{2}{2 \theta} - \Ls{2}{4 \theta} \right]
- 8 \sin (2 \theta) l_{2\theta}
\left[ 2 \Ls{3}{2 \theta} - \Ls{3}{4 \theta} \right]
\nonumber \\ && \hspace{2mm}
+ \frac{8}{3} \sin (2 \theta)
\left[ 2 \Ls{4}{2 \theta} - \Ls{4}{4 \theta} \right]
\Biggr\} 
+ {\cal O}(\ep^2) \; ,
\end{eqnarray}
where $l_{m \theta}$ and $L_{\theta}$ are defined in (\ref{Ll}).
The analytical results for the next two coefficients are expressible in terms 
of the ${\rm LsLsc}$-functions. 
In the order $\ep^2$ 
the previous combination  (\ref{combination}) 
of ${\rm LsLsc}$ functions is included.
The divergent parts of this integral were calculated 
previously in \cite{chung}.
In the regions of the variable $(M > 2m)$ the proper analytical 
continuation of all expressions can  be constructed. 
It completly coincides with the previous case $V_{mmmM}$.
The result is relatively lengthy and therefore will not be presented here. 

For the case of equal masses $M=m$ (the master-integral $D_5$) we get
\begin{eqnarray}
&& \hspace{-7mm}
(m^2)^{-1+3\ep}
(1-\ep)(1-2\ep)^2 D_5 = 
- \frac{1}{\ep^3}
- \frac{2}{3 \ep^2}
- \frac{2}{\ep} \left\{ 1 - 6 \tfrac{\Ls{2}{\tfrac{\pi}{3}}}{\sqrt{3}} \right\}
\nonumber \\ && \hspace{25mm}
- 6
+ 6 \zeta_2 \tfrac{\pi}{\sqrt{3}}
+ 6 \zeta_3 
+ 18 \tfrac{\Ls{3}{\tfrac{2 \pi}{3}}}{\sqrt{3}}
- 12 \tfrac{\Ls{2}{\tfrac{\pi}{3}}}{\sqrt{3}} \ln 3
\nonumber \\ && \hspace{-7mm}
- \ep \Biggl\{ 
 18
+ 12 \zeta_3 \tfrac{\pi}{\sqrt{3}}
+ 6 \zeta_2 \tfrac{\pi}{\sqrt{3}} \ln 3 
- \tfrac{56}{3} \zeta_3 
+ 26 \zeta_3 \ln 3 
- 63 \zeta_4 
- 6 \tfrac{\Ls{2}{\tfrac{\pi}{3}}}{\sqrt{3} } \ln^2 3 
\nonumber \\ && \hspace{0mm}
+ 18 \tfrac{\Ls{3}{\tfrac{2 \pi}{3}}}{\sqrt{3} } \ln 3 
- 12 \pi \Ls{3}{\tfrac{2\pi}{3}}
+ 18 \LS{4}{1}{\tfrac{2\pi}{3}}
+ 8 \left[ \Ls{2}{\tfrac{\pi}{3}} \right]^2
- 12 \tfrac{\Ls{4}{\tfrac{2 \pi}{3}}}{\sqrt{3} } 
\Biggr\}
\nonumber \\ && \hspace{-7mm}
+ \ep^2 \Biggl\{ 
\!-\! 54
\!+\! 3 \zeta_2 \tfrac{\pi}{\sqrt{3}} \ln^2 3 
\!+\! 12 \zeta_3 \tfrac{\pi}{\sqrt{3}} \ln 3
\!+\! 56  \zeta_3 
\!-\! 136 \zeta_4
\!-\! \tfrac{793}{18} \zeta_2 \zeta_3 
\!-\! \tfrac{3593}{9} \zeta_5
\nonumber \\ && \hspace{0mm}
\!+\! 26 \zeta_3 \ln^2 3 
\!+\! \tfrac{57}{2} \zeta_4 \tfrac{\pi}{\sqrt{3}}
\!-\! 24 \zeta_4 \ln 3 
\!-\! 2  \tfrac{\Ls{2}{\tfrac{\pi}{3}}}{\sqrt{3}} \ln^3 3 
\!+\! 9  \tfrac{\Ls{3}{\tfrac{2\pi}{3}}}{\sqrt{3}} \ln^2 3 
\!-\! 12  \tfrac{\Ls{4}{\tfrac{2\pi}{3}}}{\sqrt{3}} \ln 3
\nonumber \\ && \hspace{0mm}
\!+\! 16 \left[\Ls{2}{\tfrac{\pi}{3}} \right]^2 \ln 3 
\!+\! 6 \tfrac{\Ls{5}{\tfrac{2\pi}{3}}}{\sqrt{3}} 
\!+\! \tfrac{38}{9} \pi \zeta_2 \Ls{2}{\tfrac{\pi}{3}}
\!-\! 12 \LS{5}{1}{\frac{2\pi}{3}} 
\!+\! \tfrac{3}{2} \chi_5
\nonumber \\ && \hspace{0mm}
\!-\! 24 \pi \Ls{3}{\tfrac{2 \pi}{3}} \ln 3 
\!-\! \tfrac{46}{27} \pi \Ls{4}{\tfrac{2\pi}{3}}
\!+\! 12 \pi \Ls{4}{\tfrac{\pi}{3}}
\!+\! 36 \LS{4}{1}{\tfrac{2\pi}{3}} \ln 3 
\!-\! 48  \LsLsc{1}{1}{3}{\tfrac{\pi}{3}}
\nonumber \\ && \hspace{0mm}
\!-\! 32 \zeta_2 \ln^2 2 
\!+\! \tfrac{16}{3} \ln^4 2 
\!+\! 128 \Li{4}{\tfrac{1}{2}}
\nonumber \\ && \hspace{0mm}
  - 64 \zeta_2 \ln^3 2 
\!+\! 84 \zeta_3 \ln^2 2 
\!+\! \tfrac{64}{5} \ln^5 2 
\!+\! 384 \left[ \ln 2 \Li{4}{\tfrac{1}{2}} \ln 2 \!+\! \Li{5}{\tfrac{1}{2}} \right]
\nonumber \\ && \hspace{0mm}
\!+\! 24 \zeta_2 \ln^2 2 \ln 3 
\!-\! 42 \zeta_3 \Li{2}{\tfrac{1}{4}}
\!-\! 96 \Li{4}{\tfrac{1}{2}} \ln 3 
\!-\! 4 \ln^4 2 \ln 3 
\Biggr\} \; . 
\label{d5}
\end{eqnarray}
As a non-trivial check of these results we established full agreement
between the numerical values for the coefficients of $\ep$-expansion of 
the $D_5$-integral with the proper Laporta results \cite{laporta}.

\section{Conclusion}

In this paper, the higher order $\ep$-expansion of the diagrams shown in Fig.~(\ref{diagrams})
have been constructed (\ref{solution_v1112}), (\ref{solution_d11112}).
The coefficients of the $\ep$-expansion are parametrized in terms of generalized 
log-sin functions and ${\rm LsLsc}$-functions described in Appendix~\ref{LsLogsincos}.
At $M=m$ these integrals enter in FORM \cite{FORM} based packages 
for calculation of two-loop on-shell self-energy diagrams \cite{onshell2} $(V_{1111})$
and three-loop vacuum integrals \cite{leo,matad} $(D_5)$. The numerical values 
of the calculated coefficients (\ref{v1111_ep^2}),(\ref{v1111_ep^3}) 
and (\ref{d5}) coincide with the results presented in \cite{laporta}.

We shown that the basis of transcendental numbers for single scale diagrams 
with two- and three-massive cuts contains  new elements in addition to 
the {\it odd/even} basis of weight ${\bf 5}$ constructed in \cite{FK99,DK1}.
Some of these elements are the product of the lowest weight elements 
of the "sixth root of unity'' basis: 
\begin{eqnarray}
&& 
\zeta_2 \times \left\{\ln2 \ln3, \ln2 \ln^2 3, \Li{3}{\tfrac{1}{4}}, \Snp{1,2}{\tfrac{1}{4}}\right\} \; , 
\quad
\Li{2}{\tfrac{1}{4}} \times \left\{ \zeta_2, \zeta_3, \zeta_2 \ln 2, \zeta_2 \ln 3 \right\} \; , 
\nonumber \\ && 
\Li{4}{\tfrac{1}{2}} \times \left\{\ln 2, \ln 3 \right\} \; , 
\quad 
\ln 3 \ln^4 2.
\end{eqnarray}
Only one extra term should be added to the weight ${\bf 5}$ basis. This new constant 
can be related to a ${\rm LsLsc}_{1,1,3}$-function of the argument $\frac{\pi}{3}$.
The results of analytical calculation of three-loop master-integrals, 
evaluated in \cite{DK1} and in the present paper are available on
$$
{\tt http://theor.jinr.ru/\;\widetilde{}\;kalmykov/three\!-\!loop/master.uu}
$$. 

\noindent
{\bf Acknowledgements.}
I am grateful to A.~Davydychev, F.~Jegerlehner, and O.V.~Tarasov 
for useful discussions.
I thank Y.~Schr\"oder and A.~Vuorinen 
for correspondence concerning the typo in expression for the $\ep$-part 
of $D_4$ and for information on the results of their 
calculation before publication. 
I would like to thank S.~Mikhailov for his interest in work, and
G.Sandukovskaya for careful reading of the manuscript. 
This research was supported in part by RFBR grant \# 04-02-17192 
and the Heisenberg-Landau Programme and NATO Grant PST.CLG.980342.

\appendix

\section{Auxiliary  integrals}
\label{aint}
\setcounter{equation}{0}
Below we present the values of some integrals appearing in Sec.~2. 

\begin{eqnarray}
&& \hspace{-7mm} 
\int^\theta  d \phi  \frac{\cos \phi}{ \left( \sin \phi \right)^{3-a\ep}}
= - \frac{1}{(2-a\ep)} \frac{\left(\sin \theta \right)^{a\ep}}{\sin^2 \theta} \;, 
\\ && \hspace{-7mm} 
\Biggl(1 \!-\! \frac{a \ep}{2} \Biggr)
\int^\theta d \phi \frac{\cos \phi}{\sin^3 \phi} \left(\sin \phi \right)^{a \ep} \left(\cos \phi \right)^{b \ep}
= 
- \frac{\left(\sin \theta \right)^{a \ep} \left(\cos \theta \right)^{b \ep} }{2 \sin^2 \theta } 
- \frac{b\ep}{2} \Isc{a,b}{\theta} \; , 
\\ && \hspace{-7mm} 
\left(1 \!-\! \frac{a \ep}{2} \right)
\int^\theta d \phi \frac{\left(\sin \phi \right)^{a \ep} \left(\cos \phi \right)^{b \ep}}{\sin^3 \phi \cos \phi} 
= \left[ 1 \!-\! \frac{(a\!+\!b) \ep}{2} \right] \Isc{a,b}{\theta}
\!-\! \frac{ \left(\sin \theta \right)^{a \ep} \left(\cos \theta \right)^{b \ep} }{2 \sin^2 \theta } \;, 
\\ && \hspace{-7mm} 
(1-a\ep) \int^\theta d \phi \frac{1}{\sin^2 \phi} \left( \sin \phi \right)^{a\ep} 
\Ls{j+1}{2\phi}
= 
- \frac{\cos \theta}{ \left(\sin \theta \right)^{1-a\ep}} \Ls{j+1}{2\theta}
\nonumber \\ && \hspace{5mm}
- 2^{-a \ep} \sum_{k=0}^{\infty} \frac{(a\ep)^k}{k!} 
\left[ 
2 \frac{\ln^{k+j+1} \left( 2 \sin \theta\right) }{k\!+\!j\!+\!1}
\!+\! a \ep  \int^\theta d \phi \ln^k \left( 2 \sin \phi \right) \Ls{j+1}{2\phi} 
\right] \; , 
\\ && \hspace{-7mm} 
(1\!-\!a\ep) \int^\theta d \phi \frac{1}{\sin^2 \phi} \left( \sin \phi \right)^{a\ep} 
\Ls{j+1}{4\phi}
=
- \frac{\cos \theta}{ \left(\sin \theta \right)^{1-a\ep}} \Ls{j+1}{4 \theta}
\nonumber \\ && \hspace{5mm}
\!+\! 2^{-1-a\ep} a \ep \sum_{k=0}^\infty  
\frac{(a\ep)^k}{k!} \Ls{k+1}{2 \theta} \Ls{j+1}{4 \theta}
\nonumber \\ && \hspace{5mm}
\!+\! 2^{1-a\ep} a \ep \sum_{k=0}^\infty 
\frac{(a\ep)^k}{k!} \int^\theta d \phi  \Ls{k+1}{2 \phi}
\left[ \ln \left( 2 \sin \phi \right) \!+\! \ln \left( 2 \cos \phi \right) \right]^j
\nonumber \\ && \hspace{5mm}
\!-\! 2^{2- a\ep} \sum_{k=0}^\infty  \frac{(a\ep)^k}{k!} 
\int^\theta d \phi \frac{\cos \phi}{\sin \phi}  \ln^k \left( 2 \sin \phi \right) 
\left[ \ln \left( 2 \sin \phi \right) \!+\! \ln \left( 2 \cos \phi \right) \right]^j \;,
\end{eqnarray}
where $\Isc{a,b}{\theta}$ is defined in (\ref{Isc_ab}).
\section{Auxiliary function}
\label{LsLogsincos}
\setcounter{equation}{0}
Let us describe here some properties of the new function $\LsLsc{k}{i}{j}{\theta}$
defined as 
\begin{eqnarray}
\LsLsc{k}{i}{j}{\theta} = \int_0^\theta d \phi \Ls{k+1}{\phi}
\ln^{i-1} \left| 2 \sin  \frac{\phi}{2}  \right| 
\ln^{j-1} \left| 2 \cos  \frac{\phi}{2}  \right| \; , 
\label{LsLsc:definition}
\end{eqnarray}
where $k,i,j$ are integer numbers, $k \geq 0$ and $i,j \geq 1$,
and $\theta$ is an arbitrary real number. 
These functions have appeared in higher-order epsilon expansion of 
two- and three-loop Feynman diagrams investigated in this paper 
(see Eqs.(\ref{solution_v1112}) and (\ref{solution_d11112})). 
${\rm LsLsc}$-functions are also related to the $\ep$-expansion of 
hypergeometric functions. As was shown in \cite{DK_tokyo} 
the {\it multiple inverse binomial sums}
$
\sum_{n=1}^\infty \frac{(n!)^2}{(2n)!} \frac{z^n}{n^k} S_1(n\!-\!1) 
$
and 
$
\sum_{n=1}^\infty \frac{(n!)^2}{(2n)!} \frac{z^n}{n^k} S_1(2n\!-\!1) \;,   
$
where $S_a(n)$ is the harmonic sum, 
are expressible in terms of $\LsLsc{0}{i}{j}{\theta}$,
$\LsLsc{1}{i}{1}{\theta}$ and $\LsLsc{1}{1}{j}{\theta}$. 
Moreover, the elements $\chi_5$ and $\tilde{\chi}_5$ 
contributing to {\it odd/even} basis of a 
set of transcendental numbers  \cite{DK1} are related to
$\LsLsc{1}{3}{1}{\frac{2 \pi}{3}}$ 
and 
$\LsLsc{1}{3}{1}{\frac{\pi}{2}}$, respectively
(see Eq.~(15) in \cite{DK_tokyo}).
$\LsLsc{1}{3}{1}{\theta}$ also appears in higher 
order $\ep$-expansion of the one-loop triangle 
diagram (see Eq.~(4.5) in \cite{DK2}).
For lower values of the parameters, function (\ref{LsLsc:definition}) 
reduces to the generalized log-sine functions. In particular, 
\begin{itemize}
\item
for $k=0$ and $i=1$ or $j=1$ 
\begin{eqnarray}
\LsLsc{0}{i}{1}{\theta} & = & \LS{i+1}{1}{\theta} \;, 
\\ 
\LsLsc{0}{1}{j}{\theta} & = & \LS{j+1}{1}{\pi-\theta} -\LS{j+1}{1}{\pi} 
- \pi \left[\Ls{j}{\pi-\theta} - \Ls{j}{\pi} \right] \;. 
\end{eqnarray}
\item
for $k=0$, $i=j=1$ 
\begin{eqnarray}
\LsLsc{0}{2}{2}{\theta}
& = & 
\tfrac{1}{8} \LS{4}{1}{2\theta} - \tfrac{1}{2} \LS{4}{1}{\theta}
- \tfrac{1}{2} \left[\LS{4}{1}{\pi - \theta} - \LS{4}{1}{\pi} \right]
\nonumber \\ && 
+ \tfrac{1}{2}\pi \left[\Ls{3}{\pi - \theta} - \Ls{3}{\pi} \right] \; .
\end{eqnarray}
\item
for an arbitrary $k$ and $i=j=1$ 
\begin{eqnarray}
\LsLsc{k}{1}{1}{\theta}
= 
\theta \Ls{k+1}{\theta} 
- \LS{k+2}{1}{\theta} \; .
\end{eqnarray}
\item
for $k=1$, $i=2$ and $j=1$ 
\begin{eqnarray}
\LsLsc{1}{2}{1}{\theta}  = - \frac{1}{2} \left[\Ls{2}{\theta} \right]^2 \; .
\end{eqnarray}
\end{itemize}
For $k=1$, $i=1$ and $j=2$ we get a function, which was denoted as
$\Phi(\theta)$ in \cite{DK2} (see Eq.(2.41) in \cite{DK2}), 
\begin{eqnarray}
\LsLsc{1}{1}{2}{\theta}  = \Phi(\theta)\; .
\end{eqnarray}
The relation between the functions of the opposite arguments is 
\begin{eqnarray}
\LsLsc{k}{i}{j}{-\theta} =  \LsLsc{k}{i}{j}{\theta} \; , 
\end{eqnarray}
The symmetry relation for the new functions 
\begin{eqnarray}
\LsLsc{k}{i}{j}{\theta}
\!-\! \LsLsc{k}{i}{j}{2\pi \!-\! \theta} 
\!+\! \LsLsc{k}{i}{j}{2\pi}
= 2 \Ls{k+1}{\pi}
\left[\Lsc{i,j}{2 \pi \!-\! \theta} \!-\! \Lsc{i,j}{2 \pi} \right]  
\label{LsLsc:symmetries}
\end{eqnarray}
includes the generalized log-sine-cosine integral 
introduced in \cite{DK1} (see Appendix A.2). 
Its definition is 
\begin{eqnarray}
\Lsc{i,j}{\theta}
= 
-  \int_0^\theta d \phi 
\ln^{i-1} \left| 2 \sin  \frac{\phi}{2}  \right| 
\ln^{j-1} \left| 2 \cos  \frac{\phi}{2}  \right| \; . 
\label{Lsc:definition}
\end{eqnarray}
In particular, the following relation is valid:
\begin{eqnarray}
\int_0^\theta d \phi \Lsc{i,j}{\phi}= 
\theta \Lsc{i,j}{\theta} - \LsLsc{0}{i}{j}{\theta} \; .
\end{eqnarray}
At the same time, the relation (\ref{LsLsc:symmetries}) can be considered as definition 
of the generalized log-sine-cosine functions ($\Ls{k}{\pi}$ is a normalization constant).
For $k=0$ there is an extra symmetry relation, 
\begin{eqnarray}
\LsLsc{0}{i}{j}{\theta}
\!-\! \LsLsc{0}{j}{i}{\pi - \theta} 
\!+\! \LsLsc{0}{j}{i}{\pi}
= 
- \pi \left[\Lsc{j,i}{\pi-\theta} \!-\! \Lsc{j,i}{\pi} \right] \;. 
\label{zero}
\end{eqnarray}
When $i$ or $j$ are equal to unity, the following relations are valid:
\begin{eqnarray}
&& \hspace{-9mm}
\LsLsc{k}{i}{1}{\theta}
+ \LsLsc{i-1}{k+1}{1}{\theta}
= - \Ls{k+1}{\theta} \Ls{i}{\theta} \; , 
\\ && \hspace{-9mm}
\LsLsc{k}{1}{j}{\theta}
+ \LsLsc{j-1}{1}{k+1}{\pi-\theta}
- \LsLsc{j-1}{1}{k+1}{\pi}
=  \Ls{k+1}{\theta} \Ls{j}{\pi-\theta} \; , 
\end{eqnarray}
The values of the function $\LsLsc{k}{i}{j}{\theta}$ of the argument $\theta = 2 \pi$
can be extracted from relation (\ref{LsLsc:symmetries}). Using the symmetric properties
of  the $\Lsc{i,j}{\theta}$-function, 
\begin{eqnarray}
&& 
\Lsc{i,j}{- \theta} = - \Lsc{i,j}{\theta} \; , 
\nonumber \\ && 
\Lsc{i,j}{2\pi - \theta} = \Lsc{i,j}{2\pi} -  \Lsc{i,j}{\theta} \; , 
\nonumber \\ && 
\Lsc{i,j}{2\pi} = 2 \Lsc{i,j}{\pi} \; , \quad  \Lsc{i,j}{\pi}  =  \Lsc{j,i}{\pi} \; ,
\end{eqnarray}
it is easy to get 
\begin{eqnarray}
\LsLsc{k}{i}{j}{2 \pi} = -2 \Ls{k+1}{\pi} \Lsc{i,j}{\pi} \; ,
\label{LsLsc_2pi}
\end{eqnarray}
where the values of $\Ls{k+1}{\pi}$ and $\Lsc{i,j}{\pi}$ can be 
founded in Lewin's book \cite{Lewin}.
Taking into account that $\Ls{2}{\pi} = 0$ we get for $k=1$
\begin{eqnarray}
\LsLsc{1}{i}{j}{2 \pi} = 0 \; .
\end{eqnarray}
Using relations (\ref{LsLsc_2pi}), Eq.~(\ref{LsLsc:symmetries}) can be rewritten as 
\begin{eqnarray}
\LsLsc{k}{i}{j}{\theta} \!-\! \LsLsc{k}{i}{j}{2\pi - \theta} 
= - 2 \Ls{k+1}{\pi} \left[\Lsc{i,j}{\theta} \!-\! \Lsc{i,j}{\pi} \right] \;. 
\end{eqnarray}
From relation (\ref{zero}) we get for $\theta=\pi$
\begin{eqnarray}
\LsLsc{0}{i}{j}{\pi}
\!+\! \LsLsc{0}{j}{i}{\pi}
= \pi \Lsc{i,j}{\pi} \;. 
\end{eqnarray}
For particular values of the ${\rm LsLsc}$-functions of weight ${\bf 5}$
at $\theta=\pi,\pi/3$ and $\theta=2 \pi/3$
the PSLQ analysis \cite{PSLQ} yields 
\begin{eqnarray}
&& \hspace{-10mm}
\LsLsc{0}{3}{2}{\pi}  =
\tfrac{2}{3} \zeta_2 \ln^3 2 
\!-\! \tfrac{2}{15} \ln^5 2 
\!-\! \tfrac{7}{4} \zeta_3 \ln^2 2 
\!-\! 4 \ln 2 \Li{4}{\tfrac{1}{2}}
\!+\! \tfrac{155}{32} \zeta_5 
\!+\! \tfrac{1}{8} \zeta_2 \zeta_3 
\!-\! 4 \Li{5}{\tfrac{1}{2}}
\nonumber \\ && \hspace{25mm}
=  1.1346108755961038391083966682727 \ldots
\;,
\nonumber \\ && \hspace{-10mm}
\LsLsc{1}{1}{3}{\pi}  = 
 \tfrac{2}{5}  \ln^5 2 
\!-\! 2 \zeta_2 \ln^3 2 
\!+\! \tfrac{21}{4} \zeta_3 \ln^2 2
\!+\!  12 \ln 2 \Li{4}{\tfrac{1}{2}}
\!-\! \tfrac{7}{8} \zeta_2 \zeta_3 
\!-\! \tfrac{155}{16} \zeta_5 
\!+\!  12 \Li{5}{\tfrac{1}{2}}
\nonumber \\ && \hspace{25mm}
=  0.6301340120387385042438188494654 \ldots
\;, 
\nonumber \\ && \hspace{-10mm}
\LsLsc{1}{3}{1}{\pi}  = 
- \tfrac{2}{15} \ln^5 2 
+ \tfrac{2}{3} \zeta_2 \ln^3 2 
- \tfrac{7}{4} \zeta_3 \ln^2 2
- 4 \ln 2 \Li{4}{\tfrac{1}{2}}
+ \tfrac{155}{32} \zeta_5 
-  4 \Li{5}{\tfrac{1}{2}}
\nonumber \\ && \hspace{25mm}
=  0.8874478318089418243337609880871 \ldots \;, 
\nonumber \\ && \hspace{-10mm}
\LsLsc{1}{2}{2}{\pi}  =
 \tfrac{2}{15} \ln^5 2 
- \tfrac{2}{3} \zeta_2 \ln^3 2 
+ \tfrac{7}{4} \zeta_3 \ln^2 2 
+ 4 \ln 2 \Li{4}{\tfrac{1}{2}}
- \tfrac{279}{64} \zeta_5 
+ 4 \Li{5}{\tfrac{1}{2}}
\nonumber \\ && \hspace{25mm}
= -0.3851859504113720162670058305844 \ldots \;, 
\nonumber \\ && \hspace{-10mm}
\LsLsc{1}{3}{1}{\tfrac{\pi}{3}}  = \tfrac{11}{18} \zeta_5 
=  0.6336780725876149549802789098028 \ldots \; , 
\nonumber \\ && \hspace{-10mm}
\LsLsc{1}{3}{1}{\tfrac{2 \pi}{3}}  =
-\tfrac{55}{324} \pi\zeta_2 \Ls{2}{\tfrac{\pi}{3}}
+ \tfrac{1225}{1296} \zeta_2 \zeta_3 + \tfrac{4621}{1296} \zeta_5
+ \tfrac{23}{972} \pi \Ls{4}{\tfrac{\pi}{3}}
\nonumber \\ && \hspace{15mm}
- \tfrac{5}{18} \pi \Ls{4}{\tfrac{2 \pi}{3}}
- \tfrac{1}{3} \Ls{2}{\tfrac{\pi}{3}} \Ls{3}{\tfrac{2 \pi}{3}}
+ \tfrac{1}{3} \LS{5}{1}{\tfrac{2 \pi}{3}} - \tfrac{1}{48} \chi_5 
\nonumber \\ && \hspace{15mm}
=  0.7441484098381945153773320072924 \ldots \; ,
\nonumber \\ && \hspace{-10mm}
\LsLsc{1}{2}{2}{\tfrac{\pi}{3}}  =
- \tfrac{55}{288} \pi \zeta_2 \Ls{2}{\tfrac{\pi}{3}}
- \tfrac{3}{8} \Ls{2}{\tfrac{\pi}{3}} \Ls{3}{\tfrac{2 \pi}{3}}
+ \tfrac{23}{864} \pi \Ls{4}{\tfrac{\pi}{3}}
\nonumber \\ &&  \hspace{15mm}
- \tfrac{1}{16} \pi \Ls{4}{\tfrac{2 \pi}{3}}
+ \tfrac{361}{1152} \zeta_2 \zeta_3 
- \tfrac{41}{384} \zeta_5 
- \tfrac{3}{128} \chi_5 
\nonumber \\ && \hspace{15mm}
= -0.3438441037234077110928896695989 \ldots
\;, 
\nonumber \\ && \hspace{-10mm}
\LsLsc{1}{2}{2}{\tfrac{2 \pi}{3}}  =
\tfrac{55}{432} \pi \zeta_2 \Ls{2}{\tfrac{\pi}{3}}
+ \tfrac{1}{4} \Ls{2}{\tfrac{\pi}{3}} \Ls{3}{\tfrac{2 \pi}{3}}
- \tfrac{23}{1296} \pi \Ls{4}{\tfrac{\pi}{3}}
\nonumber \\ &&  \hspace{15mm}
+ \tfrac{5}{24} \pi \Ls{4}{\tfrac{2 \pi}{3}}
- \tfrac{1225}{1728} \zeta_2 \zeta_3 
- \tfrac{1379}{576} \zeta_5 
- \tfrac{1}{4} \LS{5}{1}{\tfrac{2 \pi}{3}}
+ \tfrac{1}{64} \chi_5 
\nonumber \\ &&  \hspace{15mm}
= -0.2676755241093223655003711724570 \ldots
\;, 
\nonumber \\ && \hspace{-10mm}
\LsLsc{0}{3}{2}{\tfrac{\pi}{3}}  =
- \LsLsc{1}{1}{3}{\tfrac{\pi}{3}}  
+ \tfrac{55}{216} \pi \zeta_2 \Ls{2}{\tfrac{\pi}{3}}
+ \tfrac{1}{2} \Ls{2}{\tfrac{\pi}{3}} \Ls{3}{\tfrac{2\pi}{3}}
+ \tfrac{223}{864} \zeta_2 \zeta_3 
\nonumber \\ &&  \hspace{15mm}
- \tfrac{793}{864} \zeta_5 
- \tfrac{79}{648} \pi \Ls{4}{\tfrac{\pi}{3}}
- \tfrac{1}{12} \pi \Ls{4}{\tfrac{2\pi}{3}}
+ \tfrac{1}{4} \LS{5}{1}{\tfrac{2\pi}{3}}
+ \tfrac{1}{32} \chi_5 
\nonumber \\ &&  \hspace{15mm}
+ 4 \Li{5}{\tfrac{1}{2}}
+ 4 \ln 2 \Li{4}{\tfrac{1}{2}} 
- \tfrac{2}{3} \zeta_2 \ln^3 2 
+ \tfrac{7}{4} \zeta_3 \ln^2 2 
+ \tfrac{2}{15} \ln^5 2 
\nonumber \\ &&  \hspace{15mm}
= -0.1728468405935728535297398287784 \ldots
\nonumber \\ && \hspace{-10mm}
\LsLsc{0}{3}{2}{\tfrac{2 \pi}{3}}  =
  \LsLsc{1}{1}{3}{\tfrac{\pi}{3}}  
- \tfrac{55}{216} \pi \zeta_2 \Ls{2}{\tfrac{\pi}{3}}
- \tfrac{1}{2} \Ls{2}{\tfrac{\pi}{3}} \Ls{3}{\tfrac{2\pi}{3}}
+ \tfrac{1565}{864} \zeta_2 \zeta_3 
\nonumber \\ &&  \hspace{15mm}
+ \tfrac{7183}{864} \zeta_5
+ \tfrac{55}{648} \pi \Ls{4}{\tfrac{\pi}{3}}
- \tfrac{1}{12} \pi \Ls{4}{\tfrac{2\pi}{3}}
- \tfrac{1}{2} \LS{5}{1}{\tfrac{2\pi}{3}}
- \tfrac{1}{32} \chi_5 
\nonumber \\ &&  \hspace{15mm}
- 12 \Li{5}{\tfrac{1}{2}}
- 12 \ln 2 \Li{4}{\tfrac{1}{2}} 
+ 2 \zeta_2 \ln^3 2 
- \tfrac{21}{4} \zeta_3 \ln^2 2 
- \tfrac{2}{5} \ln^5 2 
\nonumber \\ &&  \hspace{15mm}
= -0.2193285689662125608977695929575 \ldots
\nonumber \\ && \hspace{-10mm}
\LsLsc{1}{1}{3}{\tfrac{2 \pi}{3}}  =
- \LsLsc{1}{1}{3}{\tfrac{\pi}{3}}  
+ \tfrac{55}{216} \pi \zeta_2 \Ls{2}{\tfrac{\pi}{3}}
+ \tfrac{1}{2} \Ls{2}{\tfrac{\pi}{3}} \Ls{3}{\tfrac{2\pi}{3}}
- \tfrac{685}{864} \zeta_2 \zeta_3 
\nonumber \\ &&  \hspace{15mm}
- \tfrac{10235}{864} \zeta_5
- \tfrac{23}{648} \pi \Ls{4}{\tfrac{\pi}{3}}
- \tfrac{1}{12} \pi \Ls{4}{\tfrac{2\pi}{3}}
+ \tfrac{1}{4} \LS{5}{1}{\tfrac{2\pi}{3}}
+ \tfrac{1}{32} \chi_5 
\nonumber \\ &&  \hspace{15mm}
+ 16 \Li{5}{\tfrac{1}{2}}
+ 16 \ln 2 \Li{4}{\tfrac{1}{2}} 
- \tfrac{8}{3} \zeta_2 \ln^3 2 
+ 7 \zeta_3 \ln^2 2 
+ \tfrac{8}{15} \ln^5 2 
\nonumber \\ &&  \hspace{15mm}
=  0.4544220738685365169842148301889 \ldots
\;.
\end{eqnarray}
where the high-precision numerical evaluation of the generalized log-sine functions
was performed with the ${\rm lsjk}$-program \cite{lsjk} and 
$$
\LsLsc{1}{1}{3}{\tfrac{\pi}{3}} = 
0.32403774485559909073259711644693958993 \ldots \; .
$$

\end{document}